\documentclass{article}
\usepackage{spconf,amsmath,graphicx}
\usepackage{multirow}
\usepackage{pifont}
\usepackage{url}
\usepackage{pgfplots}
\usepackage{booktabs}
\usepackage{tabularx}
\pgfplotsset{compat=newest}
\newcommand{\cmark}{\ding{51}}
\newcommand{\xmark}{\ding{55}}

\title{Self-Supervised Learning for Speech Enhancement Through Synthesis}
\name{Bryce Irvin$^{1,2}$, Marko Stamenovic$^{1}$, Mikolaj Kegler$^{1}$, Li-Chia Yang$^{1}$}
\address{$^1$ Bose Corporation, USA, $^2$ Georgia Institute of Technology, USA}

\begin{document}
\ninept
\maketitle
\begin{abstract}
Modern speech enhancement (SE) networks typically implement noise suppression through time-frequency masking, latent representation masking, or discriminative signal prediction. In contrast, some recent works explore SE via generative speech synthesis, where the system’s output is synthesized by a neural vocoder after an inherently lossy feature-denoising step. In this paper, we propose a denoising vocoder (DeVo) approach, where a vocoder accepts noisy representations and learns to directly synthesize clean speech. We leverage rich representations from self-supervised learning (SSL) speech models to discover relevant features. We conduct a candidate search across 15 potential SSL front-ends and subsequently train our vocoder adversarially with the best SSL configuration. Additionally, we demonstrate a causal version capable of running on streaming audio with 10ms latency and minimal performance degradation. Finally, we conduct both objective evaluations and subjective listening studies to show our system improves objective metrics and outperforms an existing state-of-the-art SE model subjectively.

\end{abstract}
\section{Introduction}
\label{sec:intro}

Speech enhancement (SE), which intends to separate voice from unwanted interference, is an essential component of voice communication systems and has been a focus of research for decades. Modern SE systems are often powered by deep neural networks (DNNs) for tasks such as telephony~\cite{demucs}, speech coding~\cite{GenerativeSC}, hearing assistance~\cite{tinylstms, weightblock}, and automatic speech recognition~\cite{se_asr}. Existing DNN-based SE approaches most commonly implement the framework of supervised learning, in which the model is trained to output the clean speech signal, given noisy input. As such, these models are typically trained \textit{from scratch}; with randomly-initialized weights

On the other hand, recent advancements in self-supervised learning (SSL) have led to the emergence of powerful models pre-trained on large corpora capable of capturing transferable representations of audio~\cite{Wav2Vec2,WavLM,UniSpeechSAT,HuBERT,ModifiedCPC,BYOLA2}.
Such SSL models can be used to obtain rich representations useful for a variety of tasks. In particular, SSL models are commonly applied in the context of automatic speech recognition~\cite{Wav2Vec2,WavLM} and paralinguistics~\cite{elbanna22_interspeech,scheidwasser2022serab} as feature extractors, especially in applications whereby task-specific data is scarce. However, the utility of SSL representations has not been as thoroughly studied in the context of SE or speech synthesis.

Some works utilize the SSL models to act as deep perceptual losses~\cite{germain2018speech}, which tend to improve the performance of SE models. Meanwhile, other studies used pre-trained SSL representations as inputs into an SE system. In  \cite{Investigating_SSL}, Huang et al. evaluate 13 different upstream models on a time-frequency masking-based speech enhancement task. Following this work, Hung et al. \cite{Boosting} show that performance can be boosted by concatenating self-supervised learning representations with raw acoustic features. In these cases, the downstream model is an LSTM or BLSTM-based denoiser architecture and is responsible for generating the time-frequency masks, which are applied to the noisy input.

While using SSL representations in the context of mask-based SE can improve performance, we hypothesize that latent embeddings may be more well-suited for directly synthesizing audio~\cite{nossier2020mapping}. This is motivated by the fact that deep SSL models decompose input into a set of features not necessarily related to the short-time Fourier transform (STFT) representation, typically used in the mask-based approaches. As such, using richer SSL features to compute the STFT masks might limit the efficacy of the former in the context of SE.

Neural vocoders~\cite{choi2021neural}, such as WaveNet \cite{WaveNet} or HiFiGAN \cite{HiFiGAN_Vocoder}, have become adept at synthesizing high-quality speech from features such as Mel spectrograms. Through varying methods, these models are able to produce audio with natural-sounding phase without access to the original time-domain signal. As phase reconstruction issues have been heavily discussed in the SE literature (e.g,~\cite{phase-sensitive}) neural vocoders present an attractive solution. In fact, several works have approached speech enhancement through synthesis (SETS) in this manner \cite{Regen, HiFiGAN_Denoising}. However, these systems incorporate modules tasked with denoising speech before resynthesis. We refer to this type of approach as pre-denoising. We suggest that pre-denoising modules are a potential bottleneck for several reasons. Cascading modules are prone to compounding errors. If the pre-denoiser and vocoder are trained separately, there exists potential for a mismatch. The vocoder expects "perfect" features, and the prediction model will inevitably make mistakes, adversely affecting the resulting speech quality.

Although joint training can help avoid compounding errors, the process of pre-denoising is inherently lossy. In attempting to isolate the speech, the process may also corrupt other acoustic cues that could be helpful for speech-noise separation and speech intelligibility preservation. Pre-denoising modules also add complexity, the necessity of which has not been validated. We hypothesize neural vocoders are powerful enough to perform speech enhancement from noisy representations directly.

In this work, we explore the utility of applying pre-trained SSL representation models to SETS. Unlike the previous exploration of the use of SSL models in SE, we avoid mask-based subtractive noise suppression by directly reconstructing the clean output from the SSL model embedding of noisy input. Additionally, we avoid the pre-denoising step in other SETS systems, instead opting for a denoising vocoder (DeVo) approach. We benchmark the efficacy of existing SSL models pre-trained on large datasets (and \textit{hand-crafted} feature baselines) in SE by coupling them with trainable denoising vocoders to synthesize noise-free speech. Furthermore, we investigate the impact of training strategy on performance and generalization. We evaluate the proposed approaches using objective perceptual metrics, as well as through a subjective listening study.

\section{Method}
\label{sec:pagestyle}

We start by using SSL models as a basis to identify representation candidates directly suitable for enhancement through synthesis. The models explored include wav2vec 2.0~\cite{Wav2Vec2}, WavLM~\cite{WavLM}, UniSpeech-SAT~\cite{UniSpeechSAT}, HuBERT~\cite{HuBERT}, Modified Contrastive Predictive Coding (CPC)~\cite{ModifiedCPC}, and BYOL-A~\cite{BYOLA2}. Several of these models come in two sizes, Base and Large, and we use Base-sized models where applicable. Most models consist of a convolutional feature encoder followed by either Transformer layers or an LSTM layer. For these, we explore using both the full model and just the feature encoder. Several models also have ``+'' configurations which have been trained with a larger amount of data that is more diverse. We evaluate these models separately as candidates as well. We use the log-Mel spectrogram (LMS) with a frame length of 1024, a hop length of 160 samples, and 128 Mel bins as a \textit{hand-crafted} representation baseline. All the encoders process audio inputs sampled at 16 kHz.

\begin{figure}
\includegraphics[width=8.5cm]{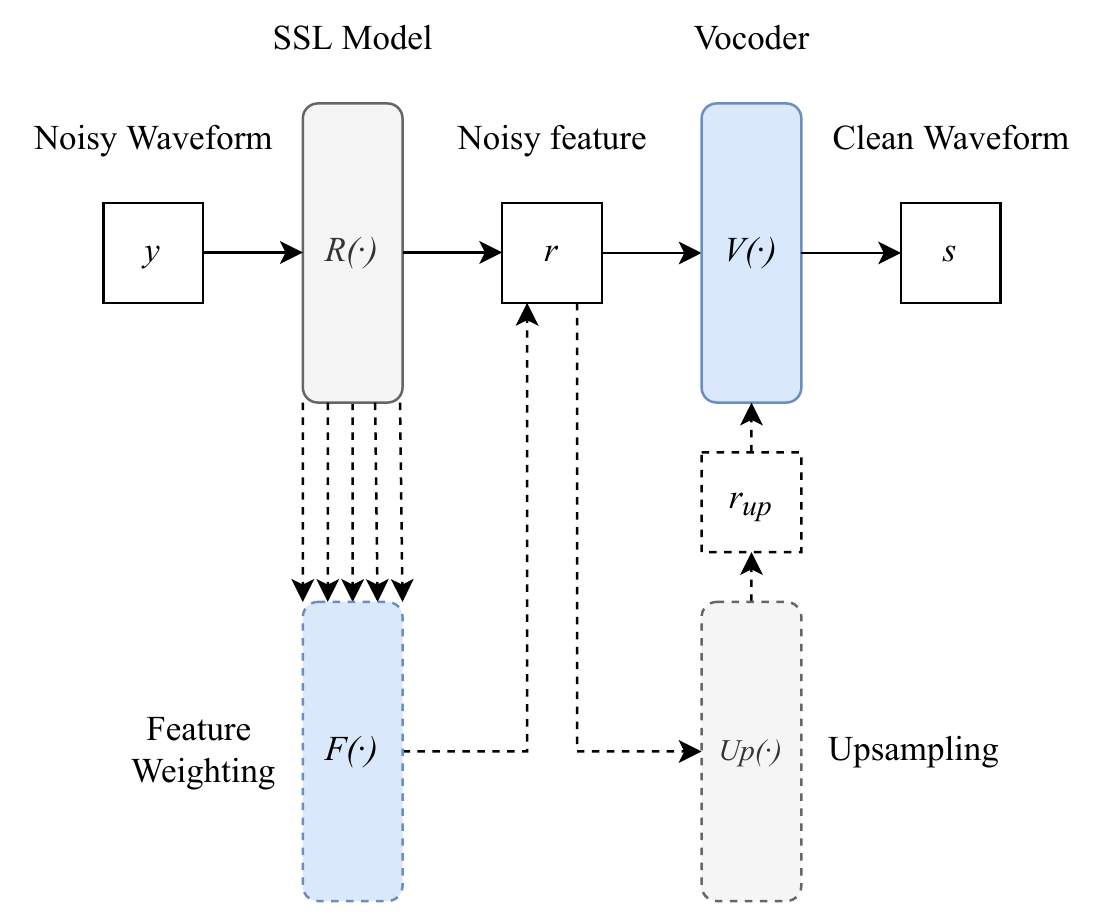}
\caption{DeVo system diagram. Components in blue are trainable and those in grey have frozen weights (unless stated otherwise). Components with dotted lines are only used for some SSL models.}
\label{fig:system}
\end{figure}

We use these models in combination with a neural vocoder, HiFiGAN~\cite{HiFiGAN_Vocoder}, to synthesize high-quality clean speech. The HiFiGAN architecture consists of transposed convolutions which progressively up-sample the input representation into a time-domain waveform and convolutions with varying receptive fields which provide the model with important context. If necessary, we up-sample representations in time using nearest neighbor interpolation to match the HiFiGAN resolution. Figure~\ref{fig:system} depicts this framework. 

Most of the surveyed SSL models contain a stack of Transformer layers following the convolutional encoder. Since previous studies have shown that the last layer is not always the most useful for a given task, we adopt the weighted sum approach following SUPERB~\cite{SUPERB}. This allows the model to emphasize or de-emphasize information from different Transformer layers. The parameters of the SSL models are fixed, and only the feature weights and vocoder parameters are optimized during training (unless stated otherwise). 

Additionally, we explore initializing HiFiGAN with pre-trained weights optimized to synthesize clean speech from the Mel spectrograms. To accept representations with different feature dimensions, the first layer of HiFiGAN must be altered. We adopt the cross-modality pre-training approach proposed by~\cite{wang2016crossmod} and average the weights of the pre-trained first layer, then use this average to initialize a new size-adapted input layer. In initial experiments, we found that this has a negligible impact on feature weighting and training stability, but the resulting audio output quality is higher. Thus, we adopt this technique for all training setups in our candidate search. 


\section{Experimental Setup}
\label{sec:typestyle}

\subsection{Dataset}
\label{ssec:subhead}
We use the dataset from the Interspeech 2020 Deep Noise Suppression (MSDNS) Challenge \cite{DNS_Challenge} for training and testing. The training dataset consists of 500+ hours of clean speech and 100+ hours of noise. We mix the speech and noise at various SNR levels, sampled from a uniform distribution between -5 and 15 dB. We employ a LUFS-based SNR calculation for more perceptually relevant mixtures and to de-emphasize the effects of impulsive noises~\cite{lufs}. The evaluation dataset consists of 150 samples of noisy speech samples and their respective clean references. Though reverberant versions of these samples are available, we do not utilize these as de-reverberation is outside of the scope of this work. In the interest of comparison to other speech enhancement work, we also report evaluation results on the Valentini et al. \cite{Valentini} dataset, which has 824 utterances mixed with noise across four SNR levels: 2.5, 7.5, 12.5, and 17.5 dB. All the audio clips are sampled at 16 kHz.

\subsection{Losses}
\label{ssec:subhead}

HiFiGAN is trained adversarially and as adversarial training can be slow and complex, we use a non-adversarial criterion for our candidate search, the Phase-Constrained Magnitude Loss (PCM) introduced in \cite{PCM}, defined as
\begin{equation}
L_{PCM}(s, \hat{s}) = \frac{1}{2}\cdot L_{SM}(s, \hat{s})+ \frac{1}{2}\cdot L_{SM}(n, \hat{n})
\end{equation}
where \begin{math}s\end{math} and \begin{math}n\end{math} are speech and noise, and \begin{math}L_{SM} \end{math} is a spectral magnitude loss defined as
\begin{multline}
L_{SM}(s,\hat{s}) = \frac{1}{T\cdot F}\sum_{t=0}^{T-1}\sum_{f=0}^{F-1}[(| S_{r}(t,f)|+|S_{i}(t,f)|) - \\ 
(|\hat{S}_{r}(t,f)| +|\hat{S}_{i}(t,f)|)] 
\end{multline}
where \begin{math}S_{r}\end{math} and \begin{math}S_{i}\end{math} are the real and imaginary components of the Short-Time Fourier Transform (STFT). We set the frame length and hop length for the STFT to 1024 and 160, respectively.

Once a suitable candidate is found, we train this candidate with HiFiGAN's original loss \cite{HiFiGAN_Vocoder}. As mentioned previously, HiFiGAN is trained adversarially, and the generator ($G$) and discriminator ($D$) objectives are defined as
\begin{gather} \mathcal{L}_{G} = \mathcal{L}_{Adv}(G;D) + \mathcal{L}_{FM} + \mathcal{L}_{Mel}(G) \\
 \mathcal{L}_{D} = \mathcal{L}_{Adv}(D;G) \end{gather}
where \begin{math} \mathcal{L}_{Adv} \end{math}, \begin{math} \mathcal{L}_{FM} \end{math}, \begin{math} \mathcal{L}_{Mel}(G) \end{math} are the adversarial loss, feature matching loss, and log-Mel spectrogram loss respectively. Our only modification is adding an L1 noise estimation term to the log-Mel spectrogram loss akin to the PCM loss in equation (1).

\subsection{Metrics}
\label{ssec:subhead}

To find the most suitable SSL front-end, we initiate our investigation by looking to Short-Time Objective Intelligibility (STOI) \cite{taal2011algorithm}  and  Perceptual Evaluation of Speech Quality (PESQ) \cite{pesq} as our metrics. While STOI is often reported in SE literature, experimentally we find that vocoder-denoised samples result in STOI degradations, even if the samples are of high subjective quality (as shown in our subjective listening study). Although \cite{DNS_Challenge} suggests metrics, such as PESQ, do not always correlate well with perceptual quality, we find it to be a coarse metric capable enough to point the exploration in the right direction. Even if raw scores do not exactly predict human perception, we find $\Delta$PESQ to be indicative of differences in model performance. $\Delta$PESQ is defined as \begin{math} PESQ(enhanced, clean) - PESQ(noisy, clean)\end{math}. In light of this, we choose to rely on $\Delta$PESQ for our candidate model selection. 

In the evaluation of our adversarially trained models, we explore a wider range of metrics: NISQAv2 \cite{mittag2021nisqa}, DNSMOS P.835 \cite{dnsmos2}, and Composite objective measures \cite{composite_metrics}. NISQAv2 and DNSMOS P.835 are deep learning-based non-intrusive perceptual quality predictors. The former predicts a single MOS score and the latter outputs a score for speech quality (SIG), background noise quality (BAK) and overall quality (OVRL). Composite scores (CSIG, CBAK, COVRL) are similar to DNSMOS P.835 but utilize regression-based analysis on a selection of objective measures in contrast to deep learning.

Since the human perception of sound is highly subjective and non-linear, perceptual assessments are still considered the ``golden standard'' in evaluating speech quality \cite{dnsmos2}. Thus, in addition to the objective evaluation, we conduct a listening study using outputs of various model variants presented to human raters. The study is based on ITU-T P.808~\cite{p808} and employs 20 randomly selected noisy samples from the MSDNS test set~\cite{DNS_Challenge}, each processed using selected approaches. Participants are asked to rate the overall quality of noisy speech samples processed using different methods using a 5-point Likert scale. The presentation order of unlabelled (i.e. double-blind) audio clips is randomized for each participant.

\begin{table}[]
\centering
\caption{Selected speech representations, their properties, and the candidate search results. \textit{NE} stands for noise exposure and indicates models that have been pre-trained using noisy speech data. \textit{CFE} indicates convolutional feature encoders obtained from a larger model. \textit{Dim.} indicates the dimensionality of the SSL feature.}
\vspace{6pt}
\label{tab:candidate_search}
\begin{tabular}{l|ccccc}
\toprule
\textbf{Representation}        & \textbf{NE} & \textbf{CFE} & \textbf{Dim.} & \textbf{Params} & \textbf{$\Delta$PESQ} \\ \hline
LMS                            & -              & -              & 128  & -               & 0.400                 \\ \hline \hline
BYOL-A                        & \cmark         & \xmark           & 3072  & 6.3M            & 0.259               \\ \hline 
\multirow{2}{*}{HuBERT}        & \xmark          & \xmark            & 768 & 95M             & 0.189               \\
                               & \xmark          & \cmark           & 512 & 4.2M            & 0.286               \\ \hline
\multirow{2}{*}{Modified CPC}  & \xmark          & \xmark            & 256 & 1.8M             & 0.448               \\
                               & \xmark          & \cmark           & 256 & 1.5M            & \textbf{0.458 }              \\ \hline
\multirow{4}{*}{UniSpeech-SAT} & \xmark          & \xmark            & 768 & 95M             & 0.174               \\
                               & \xmark          & \cmark           & 512 & 4.2M            & 0.417               \\
                               & \cmark         & \xmark            & 768 & 95M             & 0.180                \\
                               & \cmark         & \cmark           & 512 & 4.2M            & 0.187               \\ \hline
\multirow{2}{*}{Wav2Vec2}      & \xmark          & \xmark            & 768 & 95M             & 0.206               \\
                               & \xmark          & \cmark           & 512 & 4.2M            & 0.335               \\ \hline
\multirow{4}{*}{WavLM}         & \xmark          & \xmark            & 768 & 95M             & 0.232               \\
                               & \xmark          & \cmark           & 512 & 4.2M            & 0.344               \\
                               & \cmark         & \xmark            & 768 & 95M             & 0.117               \\
                               & \cmark         & \cmark           & 512 & 4.2M            & 0.234              
\end{tabular}
\end{table}

\begin{table*}[]
\centering
\caption{Results of adversarial training evaluation. \textit{FT} indicates models in which Modified CPC is jointly finetuned with the denoising vocoder. \textit{S} indicates models with causal convolutions that are streaming-compatible.}
\vspace{6pt}

\begin{tabular}{l|l|c|c|ccc|ccc|c|c}
\toprule
\multirow{2}{*}{\textbf{Dataset}} & \multirow{2}{*}{\textbf{Model}} & \multirow{2}{*}{\textbf{Causal}} & \multirow{2}{*}{\textbf{NISQAv2}} & \multicolumn{3}{c|}{\textbf{DNSMOS P.835}}    & \multicolumn{3}{c|}{\textbf{Composite}}                            & \multirow{2}{*}{\textbf{PESQ}} & \multirow{2}{*}{\textbf{STOI}} \\ \cline{5-10}
                                  &                                 &                                  &                                   & \textbf{Sig}  & \textbf{Bak}  & \textbf{Ovl}  & \textbf{Sig}  & \textbf{Bak}  & \textbf{Ovl}                       &                                &                                \\ \hline
\multirow{5}{*}{MSDNS}            & Noisy (unprocessed)                           & -                                & 2.53                              & 3.39          & 2.33          & 2.34          & 3.19          & 2.53          & \multicolumn{1}{c|}{2.35}          & 1.58                           & \textbf{0.92}                  \\
                                  & LMS                             &  \xmark                                & 4.21                              & 3.51          & 3.96          & 3.21          & 3.19          & 1.91          & \multicolumn{1}{c|}{2.37}          & 1.63                           & 0.86                           \\
                                  & Modified CPC                    &  \xmark                                & 4.32                              & 3.53          & \textbf{4.01} & 3.25          & 3.24          & 2.22          & \multicolumn{1}{c|}{2.50}          & 1.82                           & 0.87                           \\
                                  & Modified CPC FT                 &   \xmark                               & \textbf{4.34}                     & \textbf{3.56} & 3.99          & \textbf{3.26} & \textbf{3.69} & \textbf{2.60} & \multicolumn{1}{c|}{\textbf{2.96}} & \textbf{2.26}                  & 0.91                           \\
                                  & Modified CPC FT-S               &  \cmark                                & 4.19                              & 3.51          & 3.86          & 3.16          & 3.50          & 2.44          & \multicolumn{1}{c|}{2.77}          & 2.07                           & 0.90                           \\ \hline \hline
\multirow{5}{*}{Valentini}        & Noisy (unprocessed)                           & -                                & 2.85                              & 3.28          & 2.88          & 2.54          & \textbf{3.36} & 2.45          & \multicolumn{1}{c|}{2.64}          & 1.97                           & \textbf{0.92}                  \\
                                  & LMS                             & \xmark                                 & 3.77                              & 3.41          & 3.73          & 3.01          & 2.86          & 2.01          & \multicolumn{1}{c|}{2.26}          & 1.76                           & 0.86                           \\
                                  & Modified CPC                    & \xmark                                 & 3.81                              & \textbf{3.42} & \textbf{3.77} & \textbf{3.03} & 3.09          & 2.23          & \multicolumn{1}{c|}{2.47}          & 1.94                           & 0.87                           \\
                                  & Modified CPC FT                 &  \xmark                                & \textbf{3.82}                     & 3.38          & 3.63          & 2.93          & 3.35          & \textbf{2.59} & \multicolumn{1}{c|}{\textbf{2.77}} & \textbf{2.27}                  & 0.90                           \\
                                  & Modified CPC FT-S               &  \cmark                                & 3.78                              & 3.28          & 3.55          & 2.81          & 3.17          & 2.46          & \multicolumn{1}{c|}{2.59}          & 2.10                           & 0.89                          
\end{tabular}
\label{tab:adv_msdns_valentini_results}
\end{table*}

\section{Results}
\label{sec:majhead}

\subsection{Candidate Search}
\label{ssec:subhead}

Table \ref{tab:candidate_search} shows the results of the SSL candidate search. From these results, we can see that convolutional feature encoders tend to outperform models with Transformer layers. We also see that models exposed to noise during training show no clear advantage over their counterparts. In fact, in some cases the reverse is true. 

Out of these candidates, we choose to further explore the feature encoder of Modified CPC. Not only does it show the best performance in terms of $\Delta$PESQ, but it is the least computationally complex of all the candidates. As we are also interested in efficiency for the speech enhancement task, we conclude this is the optimal choice.

\subsection{Encoder Feature Weighting}
\label{sssec:subsubhead}

\begin{figure}[h]
\includegraphics[scale=0.6]{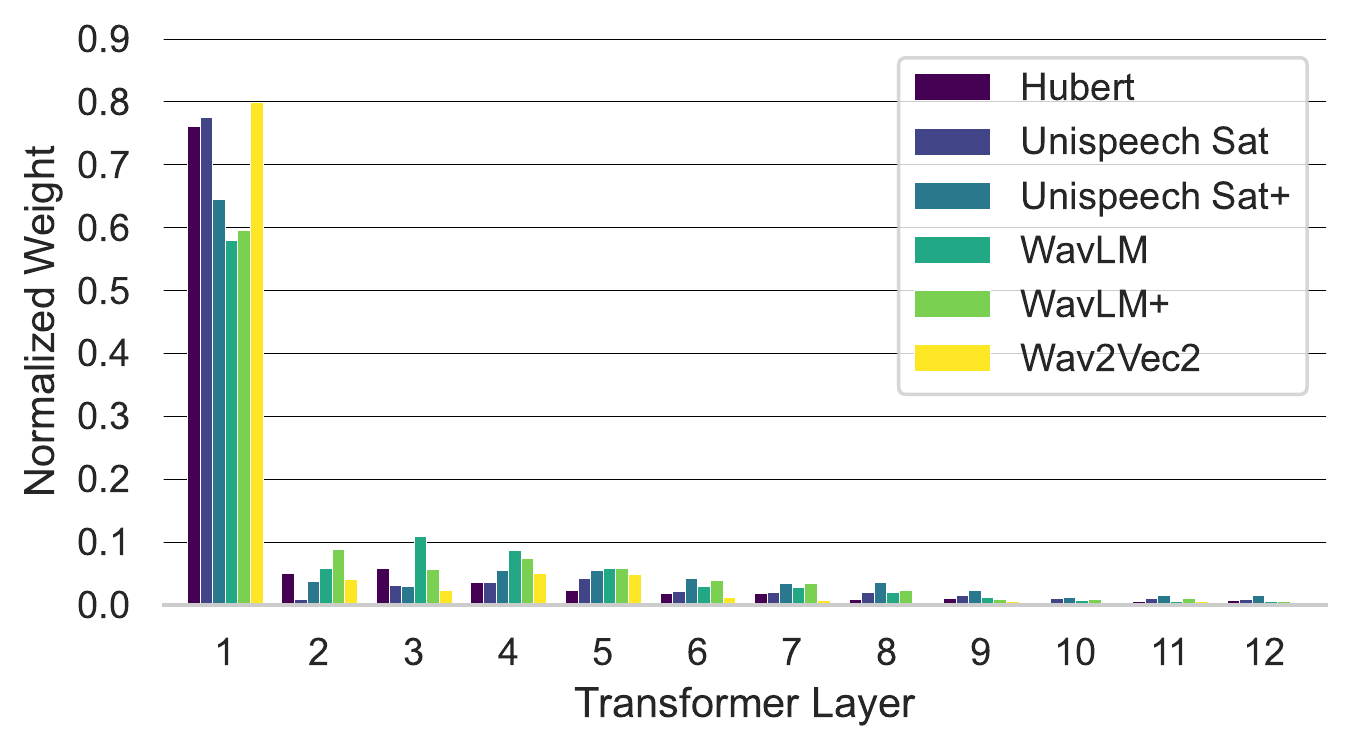}
\vspace{-12pt}
\caption{SSL transformer layer weighting. The weights have been unit-normalized for each model.}
\label{fig:Feature_Weighting}
\end{figure}

Figure~\ref{fig:Feature_Weighting} depicts the results of feature weighting for SSL models with Transformer layers. We find that in all training variations, the first layer output receives a much higher weight than all other outputs. These results are similar to the findings in~\cite{Investigating_SSL}, which suggest earlier Transformer layers contain the detailed acoustic information needed for tasks like speech enhancement, and later layers are not as critical. This is also consistent with our findings in Section~\ref{ssec:subhead}, where convolutional encoders outperform Transformer layers in combination with convolutional encoders. Thus we suggest that for speech synthesis and synthesis-based enhancement, either convolutional encoders or simply the first Transformer layer should be used.

\subsection{Adversarial Training}

As the final stage of our investigation, we adopt an adversarial training regime with the feature encoder of Modified CPC. In this stage, we evaluate the effect of fine-tuning by training two models, one with a frozen encoder (\textit{Modified CPC}) and one with a trainable encoder (\textit{Modified CPC FT}). We find that finetuning the encoder increases performance. We also evaluate the effect of making all operations in the model causal (\textit{Modified CPC FT-S}), so that the system may possibly be used in a real-time environment on streaming audio. The causal configuration of the model outputs time-domain audio in 10ms blocks with no overlap or lookahead, thus the total overall audio latency of the system is 10ms. For comparison to these SSL based results, we also adversarially train the same DeVo model configuration with log-Mel spectrogram (LMS) audio representation. Table \ref{tab:adv_msdns_valentini_results} depicts these results for the MSDNS test set, and the Valentini test set. We observe that the fine-tuned model in both cases has the strongest performance, across the vast majority of the objective metrics, so we use this version in our listening study. 

We observe that metric improvements across the board are higher for MSDNS than Valentini, which can be attributed to the overall higher SNRs of the noisy mixtures in Valentini, as compared to the MSDNS dataset. 
We also note that across both datasets, the increase in DNN-based perceptual quality predictors (DNSMOS and NISQAv2) is generally higher than that of the traditional signal processing correlation-based measures (PESQ, STOI, Composite). In fact, STOI decreases in comparison to the noisy signal. We hypothesize that this may be due to  differences between the vocoder-synthesized clean signal and the original clean signal which may be imperceptible to humans but severely deleterious to correlation-based metrics. We note that in experiments, we found that using the HiFiGAN-re-synthesized output of a clean input as the reference for STOI and PESQ resulted in significantly better scores, but a comprehensive analysis of appropriate metrics for synthesis-based enhancement is out of scope here. The incongruity between DNN-based quality predictors and traditional objective metrics motivates our subjective listening study.


\subsection{Subjective Evaluation - Listening Study}
For our perceptual listening study, we select the following conditions: unprocessed noisy speech, clean speech, DeVo using LMS features, DeVo using Modified CPC FT and Demucs\footnote{\url{https://github.com/facebookresearch/denoiser}}, a well-known waveform-to-waveform SE system~\cite{demucs} also trained on MSDNS. Table \ref{tab:listening_test_results} illustrates the averaged mean opinion scores (MOS)~\cite{p808} assigned to each method by $N = 22$ independent raters across 20 audio samples. Firstly, all of the methods improve MOS by at least 0.72 with the smallest improvement by LMS. Demucs and Modified CPC FT perform considerably better and further improve the MOS by another 0.72 and 0.94, respectively (1.44 \& 1.66 total w.r.t.~noisy). To assess the significance of these differences we conduct one-way ANOVA
with posthoc paired t-tests corrected for multiple comparisons using Benjamini-Yekutieli method~\cite{benjamini2001control}. The ANOVA indicates a significant difference between conditions ($p < 10^{-33}$) and all the pair-wise differences between methods are significant at $p < 0.013$ (corrected). Additionally, we make the audio samples from our perceptual study available for listening\footnote{\url{https://github.com/BoseCorp/devo}}.


\begin{table}[]
\centering
\caption{Results of the P808 listening study ($N=22$)}
\vspace{6pt}
\label{tab:listening_test_results}
\begin{tabular}{l|c}
\toprule
\textbf{Method} &  \textbf{MOS} \\ \hline
Noisy (unprocessed)                          & 1.86                             \\

LMS                     & 2.58                             \\

Demucs~\cite{demucs}                          & 3.30                        \\ 
Modified CPC FT                 & \textbf{3.52}                         \\\hline\hline
Clean                           & 4.48                             \\

\end{tabular}
\end{table}

\section{Conclusion}
\label{sec:print}

In this paper, we proposed that neural vocoding, when combined with a rich input feature, could be powerful enough to achieve speech enhancement without pre-denoising the speech or speech feature, a step that may limit the expressivity of the overall system. We utilized several different SSL models as a framework to explore input representations for this task and presented the results of our candidate search. We demonstrated that our synthesis-based model can run in a causal configuration with only slight degradation in objective metrics. Finally, we conducted a perceptual listening study to verify the quality of our model with human raters. Our results suggest: 1) Denoising vocoders are indeed capable of speech enhancement. 2) Earlier layers in SSL models tend to be better for the synthesis task. 3) The approach can achieve comparable or better results than other state-of-the-art speech enhancement methods.
In the future, we hope to analyze the properties of the SSL representations to reveal the reasons for their differences in performance. Additionally, we would like to explore DeVo for other speech tasks such as dereverberation and bandwidth extension.


\bibliographystyle{IEEEbib}
\bibliography{refs}

\end{document}